\documentstyle[editedvolume,epsf]{crckapb}     
\newcommand{\bef}{\begin{figure}}    
\newcommand{\eef}{\end{figure}}    
\newcommand{\be}{\begin{equation}}    
\newcommand{\ee}{\end{equation}}    
\newcommand{\etal}{{\it et al.}}    
\newcommand{\hmp}{h^{-1}Mpc}

\def\spose#1{\hbox to 0pt{#1\hss}}    
\def\ltapprox{\mathrel{\spose{\lower 3pt\hbox{$\mathchar"218$}}    
 \raise 2.0pt\hbox{$\mathchar"13C$}}}    
\def\gtapprox{\mathrel{\spose{\lower 3pt\hbox{$\mathchar"218$}}    
 \raise 2.0pt\hbox{$\mathchar"13E$}}}    
\def\inapprox{\mathrel{\spose{\lower 3pt\hbox{$\mathchar"218$}}    
 \raise 2.0pt\hbox{$\mathchar"232$}}}    
    
\begin{opening}    
\title{ON THE AVERAGE DENSITY OF GALAXIES}    
\author{M. Montuori, F. Sylos Labini and L. Pietronero, }    
\institute{
Dipartimento di Fisica, Universit\'a    
di Roma "La Sapienza",\\    
Piazzale A. Moro 2, 00185 Roma,     
Italy and INFM unit of Roma 1 }    
\end{opening}    
    
\runningtitle{ON THE MEAN DENSITY OF GALAXIES}    
    
\begin{document}    
\begin{abstract}    
    
The properties of spatial distribution of luminous matter are investigated     
 analysing all the available three dimensional catalogues of galaxies.     
In standard view, 
galaxies are believed to  
have a fractal distribution at small scale     
with a crossover to an homogeneous one at large scale.    
However up to now, the quantitative determination of this   
presumed homogeneity scale is still lacking.    
Contrary to such expectations, observational     
results show, in fact, a very inhomogeneous galaxy distribution.     
 Some years ago we criticise the standard     
statistical approach and proposed a new one     
based on the concepts and methods of modern   
statistical analysis.    
The main result of new analysis is that, contrary to the   
conclusion of standard methods,     
the distribution of galaxies in the available samples,     
does not show any crossover to homogeneity, 
but has fractal correlations   
(with dimension $D \approx 2$)     
up to the limits of present three dimensional catalogs    
($\approx 1000 h^{-1} Mpc$).    
The very first consequence of this result is that the standard approach     
is incorrect for all the length scale probed until now; moreover     
it calls for fascinating conceptual implication for the theoretical challenge     
in this field.    
    
\end{abstract}    
    
\section{Introduction}    
The question of matter homogeneity on large scale 
is a very important one,     
since it is the basic assumption of standard cosmological model.    
Redshift surveys of galaxies and clusters are naturally the main tools for  this     
study, since they allow a direct analysis of spatial properties of luminous     
matter distribution.    
Galaxy distribution is far from homogeneous on small scale and     
large scale structures (filaments and walls) appear to be limited  
only by the boundary of the     
sample in which they are detected.     
There is currently an  
acute debate on the result of the statistical     
analysis of large scale features.    
The standard interpretation assests that  
large scale homogeneity can be derived from  
isotropy  of 2D data \cite{dav97}. 
We would point out that angular homogeneity  
does not imply homogeneity  
in the corresponding 3D sets \cite{slmp97}, \cite{msl97}. 
At this point, it is more reliable  
to analyse 3D catalogs.  
In such a study there are two  
main open issues:

\begin{itemize}    
\item     
the first is whether the 3D data are reliable or not. 
Some authors \cite{dav97} 
consider many 3D catalogues too   
small to represent a fair     
statistical sample of the universe, or biased by several   
effects (non uniformity in luminosity,     
extinction from our galaxy, incompleteness, etc..). 
 
\item    
the second point regards how to  
analyse these catalogs. 
The standard approach consists of the evaluation of     
the two point correlation function $ \xi(r)$ 
\cite{dp83}. On what  
follows, we show that this is not the correct analysis  
and we introduce a new one based on  
the concepts of modern statistical mechanics.    
\end{itemize}    
    
Leaving aside these open  
controversies, it is broadly believed  
that galaxies have  
fractal distribution extending up to     
$10 h^{-1} Mpc$, with a  crossover to      
homogeneity at nearly $20 h^{-1} Mpc$.     
    
Contrary to this conclusion, the statistical analysis     
we propose, shows that the fractal structure,
 observed on smaller scales, 
extends also to distance beyond  to $20 h^{-1} Mpc$ and  
that there is  
no evidence of homogeneity  
from the available redshift samples.    
All the current 3D survey are     
consistent with each other, with  a fractal  
dimension $D \approx 2$ up to the     
sample boundaries ($\approx 1000 \hmp$).

In section 2 and 3  
we introduce the statistical methods we will employ  
later on. The     
criticism to the standard tools, i.e. the $\xi(r)$, is  
the argument of section 4.     
The results of our analysis and comparison with the standard approach      
are reported in section 5. Finally, section 6 contains  
our main conclusions.

\section{ Statistical Methods and Correlation Properties}    
    
In this section we mention the essential properties     
of fractal structures because they will be necessary for    
the correct interpretation of the statistical analysis.     
However in no way these properties are assumed    
 or used in the analysis itself.    

A fractal consists of a system in which more     
and more structures appear at smaller and     
smaller scales and the structures at small     
scales are similar to the ones at large scales.    
The first quantitative description of these forms is     
the metric dimension.    
One way to determine it, is the     
computation of mass-length    
relation. Starting from an     
 point occupied by an object     
of the distribution, we count how     
many objects $N(r)$    
("mass") are present, in average, within a volume     
of linear size  $r$ ("length")     
\cite{man83}:    
\be    
\label{l1}    
<N(r)> = B\cdot r^{D}    
\ee    
 $D$ is the fractal dimension     
and characterises in a quantitative way    
 how the system fills the space.    
The prefactor $\:B$     
depends to the lower cut-offs of the distribution; these    
are related to the smallest scale above     
which the system is self-similar     
and below which the self similarity     
is no more satisfied.    
In general we can write:    
\be    
\label{l6}    
B = \frac {N_{*}} {{r_{*}}^{D}}    
\ee    
where $r_{*}$ is this smallest scale     
and $N_{*}$ is the number of object    
up to $r_{*}$.    
For a deterministic fractal this relation is exact, while     
for a stochastic one  it is satisfied in an average sense.    
Eq.(\ref{l1}) corresponds to a average behaviour     
of $N(r)$, that is a very fluctuating     
function; a fractal is, in fact, characterised by  large    
 fluctuations and clustering at all scales.    
We stress that eq.(\ref{l1}) is completely general, i.e. it holds also     
for an homogeneous distribution, for which $D=3$.     
From eq.(\ref{l1}), we can compute the     
average density $\:<n>$ for a sample of    
 radius $\:R_{s}$ which contains a portion     
of the structure with dimension $D$.     
Assuming for simplicity a spherical volume     
($\:V(R_{s}) = (4/3)\pi R_{s}^{3}$), we have    
\be    
\label{l2}    
<n> =\frac{N(R_{s})}{V(R_{s})} = \frac{3}{4\pi } B R_{s}^{-(3-D)}    
\ee    
If the distribution is homogeneous ($D = 3$)   the    
average density is constant and independent from the sample     
volume; in the case of  a fractal, the average density     
 depends explicitly on the sample     
size $\:R_{s}$ and it is not a meaningful     
quantity. In particular, for a fractal     
the average density is a decreasing function of the sample size and     
$<n> \rightarrow 0$ for $ R_{s} \rightarrow \infty$.    
    
It is important to note that eq.(\ref{l1})     
holds from every point of the     
system, when considered as the origin.    
This feature is related to the non-analyticity of the     
distribution.    
In a fractal every observer is     
equivalent to any other one, i.e. it holds the property     
of local isotropy around any observer     
\cite {sl94}.

\section{The conditional and conditional average density}    
    
The first quantity able to analyze the spatial properties     
of point distributions is the average density.    
Coleman \& Pietronero (1992) 
introduced     
the conditional density as:    
\be    
\label{g2}    
\Gamma(r) = \frac{<n(\vec{r}+\vec{r}_{i})    
n(\vec{r}_{i})>_{i}}{<n>}     
\ee    
where the index $i$ means that the average     
is performed over the points $r_{i}$     
of the distribution.   
In other words, we consider 
spherical volumes of radius $r$ around   
each points  of the sample and we measure   
the average density of points inside them.    
Such spherical volumes have to be   
fully contained in the sample boundaries.  
$\:<n>$ is the average density of the sample;     
this normalisation does not introduce any bias even if the average    
density is sample-depth dependent,     
as in the case of fractal distributions,     
as one can see from Eq. \ref{g3}.    
The $\Gamma(r)$ (Eq. \ref{g2})    
can be computed by the following expression    
\begin{eqnarray}    
\label{g3}    
\Gamma(r)& =lim _{\Delta r \rightarrow 0}& \frac{1}{M(r)} \sum_{i=1}^{M(r)}     
\frac{1}{4 \pi r^2 \Delta r}     
\int_{r}^{r+\Delta r} n(\vec{r}_i+\vec{r'})d\vec{r'}=\nonumber\\    
& &= \frac{1}{M(r)} \sum_{i=1}^{M(r)}    
 S(r)^{-1}_{i}\frac{dN(r)_ { i}}{dr} =    
\frac{1}{M(r)} S(r)^{-1} \frac{d<N(r)>}{dr} \nonumber\\     
& &= \frac{DB}{4 \pi}  r^{3-D}    
\end{eqnarray}    
where  $S(r)$ is the area of a spherical     
shell of radius $r$, $M(r)$ is the number of spheres of   
radius $r$ and $N(r)_{i}$ is the     
number of points in the sphere of radius $r$   
centered on the $i_{th}$ point. 
$\Gamma(r)$ is a smooth function away from the     
lower and upper cutoffs of the distribution ($r_{*}$ and     
the dimension of the sample).    
From Eq.(\ref{g3}), we can see that     
$\Gamma(r)$ is independent from the     
sample size, depending only by     
the intrinsic quantities of the distribution     
($B$ and $D$). 
If the sample is homogeneous,$D=3$,     
$\:\Gamma(r) =(DB)/(4 \pi)  = (3B)/(4 \pi) = (N_{*})/(4 \pi r_{*}/3)$     
and    
then is constant.    
If the sample is fractal, then 
$D < 3$, $\:\gamma > 0$ and     
$\Gamma(r)$ is a {\it power law}.    
For a more complete discussion we refer the reader to     
\cite{cp92}, \cite{slmp97}.    
If the distribution is fractal up     
to a certain distance $\lambda_0$,    
and then it becomes homogeneous,    
we have that:    
$$    
\Gamma(r) = \frac{BD}{4 \pi} r^{D-3} \;  r < \lambda_0    
$$    
\be    
\label{e327b}     
\Gamma(r)= \frac{BD}{4 \pi} \lambda_0^{D-3} \; r \geq \lambda_0    
\ee

It is also very useful to use the {\it conditional average density}    
defined as:     
\be    
\label{g4}    
\Gamma^*(r) = \frac{3}{4 \pi r^3} \int_{0}^{r} 4 \pi r'^2 \Gamma(r') dr'    
\ee    
This function produce an artificial smoothing of     
$\Gamma(r)$ function,     
but it correctly     
reproduces global properties \cite{cp92}.

Given a certain sample of solid angle  
$\Omega$ and depth $R_d$,    
it is important to define which is     
 the maximum distance up to which it     
is possible to compute the correlation function ($\Gamma(r)$ or $\xi(r)$).     
 We have limited our analysis to an    
effective  depth    
$R_{s}$ that is of the order of the radius of the maximum    
sphere fully contained in the sample volume \cite{cp92}.    
    
The reason why    
$\Gamma(r)$ (or $\xi(r)$) cannot    
be computed for $r > R_{s}$    
is essentially the following.    
 When one evaluates the correlation    
function beyond $R_{s}$,    
then one  makes explicit assumptions on what    
lies beyond the sample's boundary.  In fact, even in absence of    
corrections for selection effects, one    
is forced to consider incomplete shells    
calculating $\Gamma(r)$ for $r>R_{s}$,    
thereby    
implicitly assuming that what one  does not see in the part of the    
shell not included in the sample is equal to what is inside (or other    
similar weighting schemes)\cite{slmp97}.

\section{Standard analysis}  
  
At this point it is instructive to   
consider the behaviour of the standard   
correlation function $\xi(r)$.  
Coleman \& Pietronero (1992) clarify some crucial   
points of the  
such an analysis, and in particular they discuss the meaning  
of the so-called {\it "correlation length"}  
  $\:r_{0}$  
found with the standard  
approach (Davis \& Peebles, 1983; Peebles, 1993) 
and defined by the relation:  
\be  
\label{x1}  
\xi(r_{0})= 1  
\ee  
where  
\be  
\label{x2}  
\xi(r) = \frac{<n(\vec{r_{i}})n(\vec{r_{i}}  
+ \vec{r})>_{i}}{<n>^{2}}-1  
\ee  
is the two point correlation function used in the   
standard analysis.  
If the average density is not a well defined intrinsic   
property of the system, the analysis with  
$\xi(r)$ gives spurious results.  
In particular, if the system has   
fractal correlations, the average density is simply related   
to the sample size as shown by Eq.(\ref{l2}).  
In other words, it is meaningless to   
define the correlation length   
of the distribution by comparing the average   
correlation $<n(\vec{r_{i}})n(\vec{r_{i}}  
+ \vec{r})>_{i}$ to the average density   
of the sample $<n>^2$, if   
the latter depends on the sample volume.  
Following \cite{cp92}, the expression of the   
$\:\xi(r)$ for a 
fractal distribution, is:  
\be  
\label{x3}  
\xi(r) = ((3-\gamma)/3)(r/R_{s})^{-\gamma} -1  
\ee  
where $\:R_{s}$ (the effective sample   
radius) is the radius of   
the spherical volume where one computes the  
average density from Eq. (\ref{l2}).  
From Eq. (\ref{x3}) it follows that:  
  
i.) the so-called correlation  
length $\:r_{0}$ (defined as $\:\xi(r_{0}) = 1$)  
is a linear function of the sample size $\:R_{s}$  
\be  
\label{x4}  
r_{0} = ((3-\gamma)/6)^{\frac{1}{\gamma}}R_{s}  
\ee  
and hence it is a quantity without any correlation 
meaning, but it is  
simply related to the sample size.  
  
ii.) the amplitude of the $\xi(r)$ is:  
\be   
\label  {x5}  
A(R_{s}) = ((3-\gamma)/3)R_{s}^{\gamma}   
\ee

iii.) $\:\xi(r)$ is a power law only for   
\be  
\label{x6}  
((3-\gamma)/3)(r/R_{s})^{-\gamma}  >> 1  
\ee  
hence for $\: r \ltapprox r_{0}$: for larger distances  
there is a clear deviation from a  
power law behavior due to the definition of $\:\xi(r)$.  
This deviation, however, is just due to the size of  
 the observational sample and does not correspond to any real change  
of the correlation properties. It is clear that if one estimates the  
 exponent of $\xi(r)$ at distances $r \gtapprox r_0$, one  
 systematically obtains a higher value of the correlation exponent  
 due to the break of $\xi(r)$ in the log-log plot.   
This is actually the case for the analyses performed so far:  
 in fact, usually,   
$\xi(r)$ is fitted with a power law in the   
range $ 0.5 r_{0} \ltapprox    
r \ltapprox 2 r_{0}$, where we get an higher value of   
 the correlation  exponent. In particular,    
the usual estimation of this exponent  
by the  $\xi(r)$ function leads to $\gamma \approx 1.7$, different   
from $\gamma \approx 1$ (corresponding  
to $D \approx 2$), that we found by means of    
the $\Gamma(r)$ analysis \cite{slmp97}.

\section{Average density of galaxies}  
  
Here we report the measure the average density   
of galaxies in all the   
three dimensional catalogs avalaible. Our analysis   
is performed in Volume Limited samples \cite{dp83}; they obsiouvly   
contain fewer galaxies with respect the magnitude limited sample,   
but their statistical analysis is   
straigthforward and free of any assumptions.  
The main data of our correlation analysis   
are collected in Fig.\ref{fig1}   
 in which we report the   
{\it galaxy density as a function of scale}  
 for the various catalogues.   
We show two different   
kind of measure of the density of   
galaxies, {\it the conditional average density}  
and {\it the radial density}.  
The former is $\Gamma^*(r)$, i.e. the computation   
of the {\it average } density of galaxies.  
In this case, we have an average quantity and then a reliable 
statistical result. However, we have to stop our analysis to 
a scale which is in general smaller than the depth of the sample 
(see Sect.3). For the present catalogues this scale is  
$\approx 150 \hmp$. 
At larger distances we can measure the 
{\it radial density}. 
In this case, the density at scale $r$ 
is given by the number of 
galaxies up   
to distance $r$ from the Earth, divided
 for the corresponding volume.  
Of course, this measure 
can be prformed up to the whole depth of 
the sample. Since it is not an average quantity 
it is more noisy than the $\Gamma^*(r)$ and then 
less statistically reliable \cite{mslgap97}.
(For the clarity's sake, we have reported the 
radial density    
only for the deepest catalogues i.e.$ESP$ and $LCRS$).  
  
In the insert we show the   
schematic behaviour of radial density.  
At small scale, where the number of galaxies   
is low, the radial density is  
dominated by poissonian noise.  
At larger scale, where the statistics   
is large, we get the right behaviour. 
If the distribution is a fractal, the radial 
density has a power law decay as function of the scale 
like the $\Gamma^*(r)$.
   
 The results of the standard analysis   
for the same galaxy catalogues are shown in   
Fig.\ref{fig4}.   
Here we report the estimate of  
$\xi ( r )$ for the 
various volume limited samples  
of Fig. \ref{fig1}.   
The various data have different correlation length  
and then appear to be in strong disagreement with   
the each other. This is due to the fact that the   
usual analysis looks at the data from the   
perspective of  analyticity and large scale homogeneity   
(within each sample). These properties are never tested and they   
are actually   
not present in the real galaxy distributions, so the result is   
rather confusing (Fig.\ref{fig4}).  
Once the same data are analyzed within a broader   
perspective the situation becomes clear (Fig.\ref{fig1}) and the data   
of different catalogues result in agreement with each other.  
In addition in the insert of Fig.\ref{fig4} we show the dependence   
of $r_0$ on $R_s$ for all the catalogs. The linear behaviour is a consequence   
of the correlation properties of Fig.\ref{fig1} and it provides   
an additional evidence of fractal behaviour to all scales.  
 
\section{conclusion}  
  
In summary our conclusions are:  
\begin{itemize}  
  
\item  
 
The properties derived from different   
catalogues show a {\it power law   
decay} of the conditional density  
($\Gamma( r )^*$)as function of   
the scale, from $1 \hmp$ to $150 \hmp$,  
without any tendency towards homogenization (flattening).  
The same scaling behaviour with  
the same amplitude  
is found at larger scales with  
the measure of {\it radial density}.  
In addition essentially all the catalogues show   
well defined fractal   
correlations up to their limits, with the fractal dimension 
$D \simeq 2$.   
  
\item  
{\it all  the samples   
analysed are statistically rather good }  
and their properties are in agreement with each other.   
The relative position of the various lines is not arbitrary but it is fixed   
by the luminosity function, a part for the cases of   
IRAS and SSRS1 for which this is   
not possible  
This whole agreement gives a   
new perspective because, using the standard methods of analysis, the   
properties of different samples appear contradictory with each other   
and often this is considered to be a problem of the data (unfair   
samples) while, we show that this    is due to the inappropriate methods   
of analysis.  
  
\item  
These results  
imply necessarily that the value of $r_0$ (derived from the $\xi(r)$   
approach) will scale with the sample size $R_s$ as shown also from the   
specific analysis of the various catalogues \cite{slmp97}. The   
behaviour   
observed  corresponds to a fractal structure with dimension $D \simeq   
2$.

\bef  
\epsfxsize 10cm  
\centerline{\epsfbox{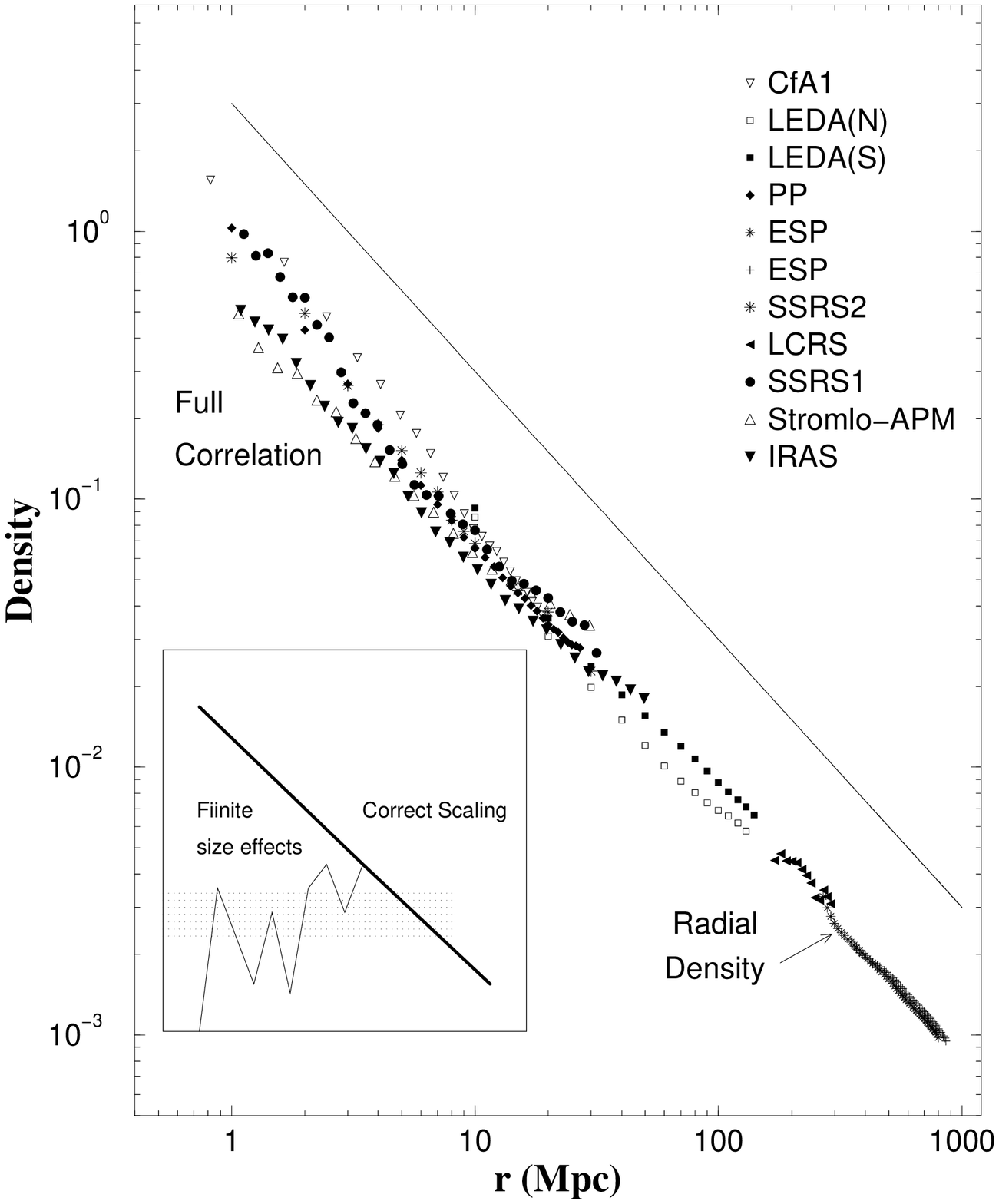}}  
\caption{\label{fig1}  Full correlation for the various available  
redshift  catalogues in the range of distances  
$0.1 \div 1000  \hmp$. A reference line with a slope   
$-1$ is also shown (i.e. fractal dimension $D = 2$).  
Up to $\sim 150 \hmp$ the density is computed by the full correlation   
analysis, while above $\sim 150 \hmp$ it is computed through the   
radial density.  
For the full correlation the data of the various catalogues are normalized  
with the luminosity function and they match very well   
with each other. This is an important test of the statistical validity and   
consistency of the various data.  
In the {\it insert panel} it is shown the   
schematic behavior of the radial   
density versus distance computed from the vertex (see text).  
The behaviour of the radial density allows us to extend the power law  
correlation up to $\sim 1000 \hmp$. However a rescaling   
is necessary to match the radial density to the   
conditional density.  
}  
\eef  
\bef  
\epsfxsize 10cm  
\centerline{\epsfbox{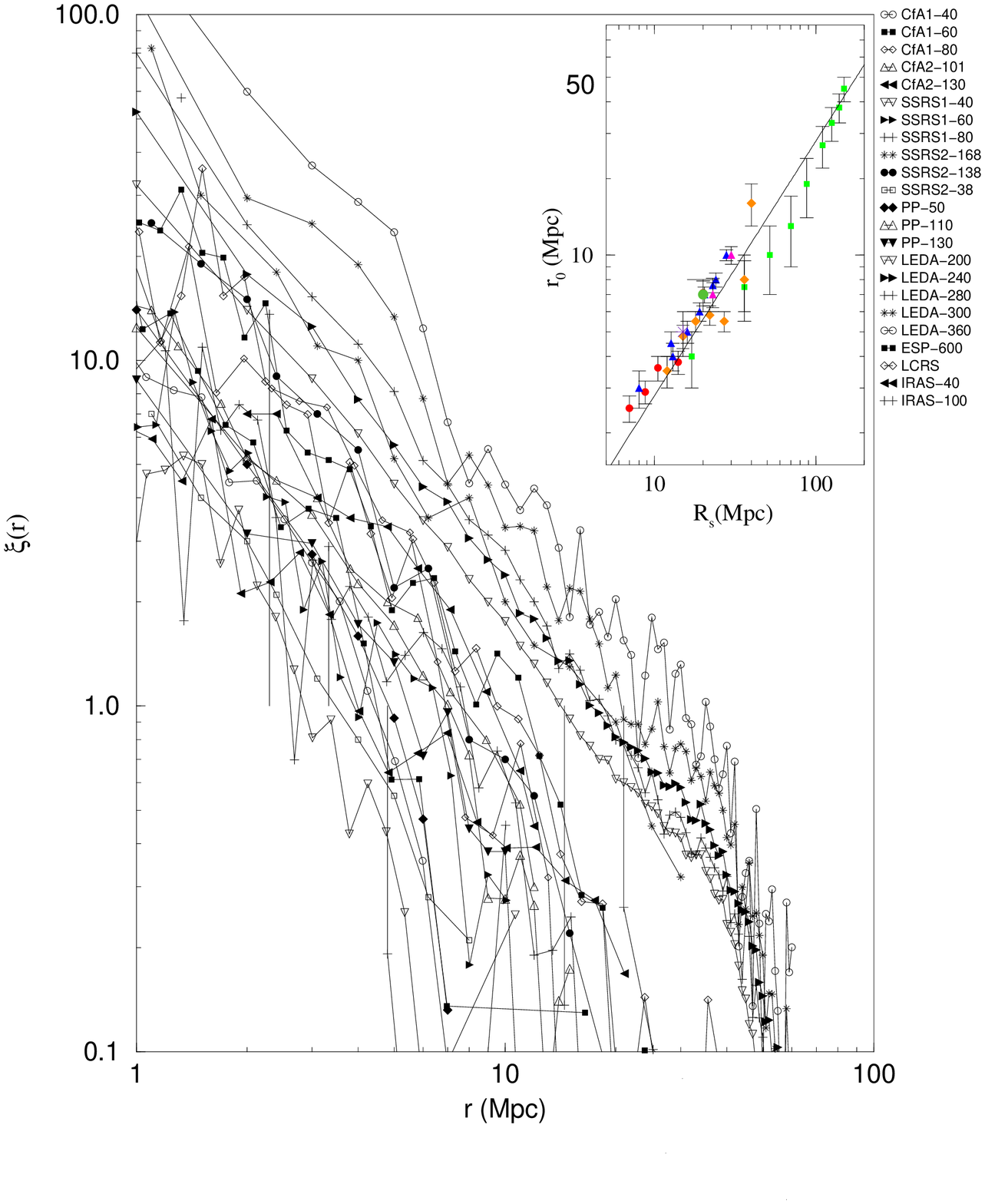}}  
\caption{\label{fig4}   
Usual analysis based on the   
function $\xi(r)$ of the same galaxy catalogues of Fig.1. This   
analysis is based on the   
a priori and untested assumption   
of the analyticity and homogeneity.  
These properties are not present in the   
real galaxy distributions and the results appear   
therefore rather confusing. This lead to the   
impression that galaxy catalogues are not   
good enough and to a variety of theoretical   
problems like the galaxy cluster   
mismatch, luminosity   
segregation, the linear and non linear evolution, etc..   
The situation changes completely and it becomes rather   
clear if one adopts   
the more general framework that is at the basis of Fig.1.   
In the {\it insert panel} we show the dependence   
of $r_0$ on $R_s$ for all the catalogs.  
The linear behaviour is a consequence   
of the fractal nature of galaxy distribution in these samples.  
}  
\eef

\item  
A possible explanation of the shift of $r_0$ is based  
on the luminosity segregation effect  
\cite{dav88} \cite{par94} \cite{ben96}.  
The fact that the giant galaxies are more  
clustered than the dwarf ones,  
i.e. that they are located in the peaks of the density field,  
has given rise to the proposition  
that larger objects may correlate up to larger  
length scales and that the amplitude of the $\:\xi(r)$ is larger  
for giants than for dwarfs one. The deeper VL subsamples  
contain galaxies that are in average  
brighter than those in the VL subsamples with  
smaller depths. As the brighter galaxies should have  
a larger correlation length the shift of $r_0$ with sample size   
could be related, at least partially,  
with  the phenomenon of luminosity segregation.  
The insert of Fig.\ref{fig4} show clearly the linear dependence  
of $r_0$ on $R_s$, which completely consistent with  
power law decay of $\Gamma ( r )$. 
In this respect, the proposed luminosity bias effect   
appears essentially irrelevant.

\end{itemize}

\section*{Acknowledgments}  
We thank for useful discussions, suggestions and collaborations  
L.Amendola, A. Amici, Yu.V. Baryshev, H. Di Nella, R. Durrer, A. Gabrielli and M. Munoz.


\begin{thebibliography}{} 
    
    
\bibitem[\protect\citeauthoryear{Benoist {\it et al.}}{1996}]{ben96} Benoist C. \etal (1996)  {\it Astrophys. J.}, in print    
    
\bibitem [\protect\citeauthoryear{Coleman and Pietronero}{1992}]{cp92}    
 Coleman, P.H. and Pietronero, L., (1992) {\it Phys.Rep.} {\bf 231},    
pp.311-391     

\bibitem[\protect\citeauthoryear{Da Costa {\it et al.}}{1994}]{dac94} Da Costa L.N., \etal (1994) {\it Astrophys. J.} {\bf 424} L1-L4    
     
 \bibitem[\protect\citeauthoryear{Davis and Peebles}{1983}]{dp83} Davis, M., Peebles, P. J. E. (1983)    
  {\it Astrophys. J.} {\bf 267}, 465-482    
    
    
\bibitem[\protect\citeauthoryear{Davis {\it et al.}}{1988}]{dav88} Davis M. \etal, (1988) {\it Astrophys. J.} {\bf  333}, L9-L12    
    
    
\bibitem[\protect\citeauthoryear{Davis }{1997}]{dav97} Davis M.,  (1997)(astro-ph/9610149)    
{\it in the Proc. of the Conference "Critical Dialogues in Cosmology"}    
N. Turok ed.     
        
\bibitem[\protect\citeauthoryear{Mandelbrot}{1982}]    
 {man83}Mandelbrot, B. (1982) {\it The Fractal Geometry of Nature,}    
Freeman, New York    

\bibitem[\protect\citeauthoryear{Montuori {\it et al.}}{1997}] {mslgap97} Montuori M., Sylos Labini, F.,    
Gabrielli, A., Amici A., Pietronero,L.  (1997)  {\it Europhys. Lett.}    
{\bf 39} 103-108    
    
    
\bibitem[\protect\citeauthoryear{Montuori and Sylos Labini}{1997}]    
{msl97}Montuori M. \& Sylos Labini F.(1997)     
{\it Astrophys. J.} in print        

\bibitem[\protect\citeauthoryear{Park {\it et al.}}{1994}]{par94} Park, C., Vogeley, M.S., Geller, M., Huchra, J.    
 (1994)    
{\it Astrophys. J.} {\bf  431} 569    
    
\bibitem[\protect\citeauthoryear{Peebles}{1993}]{pee93} Peebles, P.E.J.,     
(1993) {\it Principles of physical    
Cosmology} Princeton Univ.Press.    
    
        
    
\bibitem[\protect\citeauthoryear{Sylos Labini}{1994}] {sl94} Sylos Labini, F., (1994),  {\it Astrophys. J.}, {\bf 433}, 464-467    
    

\bibitem[\protect\citeauthoryear{Sylos Labini {\it et al.}}{1997}] {slmp97} Sylos Labini, F., Montuori, M.and  Pietronero, L.  (1996), {\it Phys. Rep.} in print    
    
\end{thebibliography}
\end{document}